\documentclass[12pt,a4paper]{iopart}

\usepackage{iopams}
\usepackage{graphicx}

\newcommand{\sgn}{\mathrm{sgn}\,}

\begin{document}

\title[Lowering the error floor of Gallager codes]{Lowering the error floor of Gallager codes: a~statistical-mechanical view}

\author{Marco Pretti}

\address{Consiglio Nazionale delle Ricerche -- Istituto dei Sistemi Complessi (CNR--ISC)
@~Dipartimento di Scienza Applicata e Tecnologia (DISAT),
Politecnico di Torino, Corso Duca degli Abruzzi 24, I-10129
Torino, Italy \\ E-mail address:
\texttt{marco.pretti@polito.it}}

\begin{abstract}

The problem of error correction for Gallager's low-density
parity-check codes is famously equivalent to that of computing
marginal Boltzmann probabilities for an Ising-like model with
multispin interactions in a non-uniform magnetic field. Since
the graph of interactions is locally a tree, the solution is
very well approximated by a generalized mean-field
(Bethe--Peierls) approximation. Belief propagation (BP) and
similar iterative algorithms are an efficient way to perform
the calculation, but they sometimes fail to converge, or
converge to non-codewords, giving rise to a non-negligible
residual error probability (error floor). On the other hand,
provably-convergent algorithms are far too complex to be
implemented in a real decoder. In this work we consider the
application of the \emph{probability-damping} technique, which
can be regarded either as a variant of BP, from which it
retains the property of low complexity, or as an approximation
of a provably-convergent algorithm, from which it is expected
to inherit better convergence properties. We investigate the
algorithm behaviour on a real instance of Gallager code, and
compare the results with state-of-the-art algorithms.

\end{abstract}

\maketitle

\section{Introduction}
\label{sec:introduction}

In recent years, new overlaps have emerged between statistical
mechanics and other fields of scientific and technical
research. In this framework, an important role has been played
by \emph{belief propagation} (BP), that is, a class of
\emph{message-passing} algorithms, suited to solving several
types of constraint-satisfaction and statistical-inference
problems~\cite{MezardMontanari2009,Yedidia2011}. It is
straightforward to show that such problems, usually defined via
graphical models, can generally be mapped onto proper
thermodynamic models of the Ising or Potts type, and ultimately
amount to computing marginals of the Boltzmann distribution of
the
latter~\cite{MezardMontanari2009,Yedidia2011,Yedidia2001,Pelizzola2005}.
A very important problem of this kind is that of error
correction for digital transmission over a noisy channel
(\emph{channel decoding}). Here we shall focus in particular on
the so-called \emph{low-density parity-check} (LDPC) codes,
which exhibit extremely good (i.e. capacity-approaching)
performance in the limit of infinite block length, and which
have therefore attracted great attention from communication
engineers~\cite{RichardsonUrbanke2003}, albeit quite long after
Gallager's original proposal~\cite{Gallager1962}. Indeed, BP as
an efficient, though suboptimal, iterative decoding algorithm
was first proposed (and simply called \emph{probabilistic
decoding}) by Gallager himself, who also realized that its
underlying approximation breaks down when the graph describing
the code structure includes cycles. In the
statistical-mechanical analogue, the inference problem arising
from Gallager codes can be viewed as an Ising model with
multispin interactions in a non-uniform magnetic
field~\cite{KabashimaSaad2004} (the infinite block-length limit
corresponding to the thermodynamic limit), while BP equations
turn out to be equivalent to the self-consistency equations
derived by the Bethe--Peierls
approximation~\cite{Bethe1935,Peierls1936}. The latter,
well-known in statistical mechanics~\cite{Burley1972}, is a
refined mean-field theory, aimed at improving the simpler
(Bragg--Williams) mean-field approximation. Note that the
Bethe--Peierls approximation dates back to 1935, but the
equivalence with BP is a relatively recent result: it was
suggested in 1998 by Kabashima and
Saad~\cite{KabashimaSaad1998}, and was proved in a rather
general setting a few years later by Yedidia, Freeman and
Weiss~\cite{YedidiaFreemanWeiss2001}. Even though the
Bethe--Peierls approximation is rigorously exact only on tree
graphs (i.e. graphs without cycles), it was found to be
extremely effective even on ``loopy''
graphs~\cite{FreyMackay1998}, in particular when the latter are
sufficiently sparse and random (and therefore locally
tree-like), as is the case for Gallager codes.

A major drawback of the BP algorithm is its possible lack of
convergence. Such a behaviour is sometimes a signal of
intrinsic inadequacies of the Bethe--Peierls approximation,
which therefore deserves further
improvements~\cite{Pelizzola2005,YedidiaFreemanWeiss2001}. If
this is not the case, it may be useful to devise strategies to
overcome convergence difficulties, and, in order to do so, one
can exploit the fact that the Bethe--Peierls approximation
admits a variational formulation (usually regarded as an
instance of Kikuchi's cluster-variation
method)~\cite{Pelizzola2005,An1988}. The variational
formulation
states~\cite{MezardMontanari2009,Yedidia2011,Yedidia2001,Pelizzola2005,YedidiaFreemanWeiss2001}
that BP fixed points are stationary points of a variational
functional, called \emph{Bethe free energy} (more precisely, it
is known that \emph{stable} fixed points of BP are local
\emph{minima} of the Bethe free energy~\cite{Heskes2003}). As a
consequence, the problem can be addressed by methods that are
able to minimize the Bethe free energy. One such method has
been proposed by Heskes, Albers and Kappen in the form of a
double-loop algorithm, performing a sequence of partial
optimizations, which can be proved to converge to a local
minimum~\cite{HeskesAlbersKappen2003}. Unfortunately, a
double-loop strategy is often practically unfeasible, because
of the very long computation time required. For this reason, in
a previous paper we proposed a single-loop iterative algorithm,
which can be regarded either as an approximation of the
aforementioned double-loop algorithm, or as a modified BP
algorithm with over-relaxed dynamics, and from there it is
denoted as \emph{probability-damping belief propagation}
(PDBP)~\cite{Pretti2005}. The latter algorithm (actually, a set
of algorithms, specified by an adjustable \emph{damping
parameter}) can be verified empirically to be more stable and
faster than ordinary
BP~\cite{Pretti2005,SomDattaSrinidhiChockalingamRajan2011,WangLiGaoWang2013},
with just a slightly larger numerical complexity. Note that the
interpretation of PDBP as an approximation of the double-loop
algorithm is relevant to get a qualitative understanding of how
it works and also, more practically, to drive the choice of the
damping parameter value.

In this paper, we shall investigate the behaviour of PDBP (and
a further variant) as a decoding algorithm on a real instance
of Gallager code. The motivation behind this analysis is that
the performance of real (finite length) Gallager codes under BP
decoding (and other iterative algorithms as well) turns out to
be significantly worse than expected, with the onset of a
non-negligible residual error probability, even in the
practically relevant regime of high signal-to-noise
ratio~\cite{MacKayPostol2003,Richardson2003,StepanovChernyakChertkovVasic2005}.
Such a phenomenon, called \emph{error-floor} in the jargon of
coding theorists, is mostly ascribed to specific configurations
of the channel errors (called \emph{trapping sets} or the
like), which hamper the convergence of the decoding algorithm.
Due to the technical importance of the subject, great efforts
in research have been made to solve this problem, following
mainly two different routes, namely, either eliminating
trapping sets by changing the graphical structure of the code
(usually removing short
cycles)~\cite{HuEleftheriouArnold2005,AsvadiBanihashemiAhmadian-Attari2011},
or modifying the algorithm dynamics, in order to make it less
sensitive to trapping
sets~\cite{HanRyan2009,YedidiaWangDraper2011}. Along the latter
line, considered in this paper, several algorithms have been
proposed, among which a noticeable case is the
\emph{difference-map belief propagation} (DMBP) algorithm by
Yedidia, Wang and Draper~\cite{YedidiaWangDraper2011}. Since
this algorithm exhibits significant performance improvement
with respect to standard BP approaches, with a moderate
increase in complexity, we shall use it as a benchmark for
evaluating our algorithms.

The paper is organized as follows. In section~\ref{sec:analogy}
we briefly review the formal analogy between a Gallager code
and a multispin Ising model. As a model of a noisy channel, we
consider the simple \emph{binary symmetric channel} (BSC), even
though the analogy holds even for more realistic models such as
the \emph{additive white Gaussian noise} (AWGN) channel. In
section~\ref{sec:algorithms} we introduce the BP algorithm and
its variants for the model of interest, using the specific
formulation suitable for binary variables and usually denoted
as \emph{effective fields} or \emph{log-likelihood ratios} (in
statistical mechanics or coding theory, respectively).
Section~\ref{sec:results} is the central section, where all the
results of our simulations are illustrated and discussed.
Section~\ref{sec:conclusions} contains some concluding remarks.

\section{Gallager codes and multispin Ising model}
\label{sec:analogy}

Let ${\bi{s} \equiv [s_1,\dots,s_N]}$ denote an encoded
sequence of binary digits (bits), transmitted over a BSC.
According to a parity-check coding scheme, such a sequence
cannot be any possible sequence of $N$ bits, but it must indeed
satisfy some prescribed parity checks. A parity check is
defined by a (usually small for LDPC) subset of indices ${a
\subseteq \{1,\dots,N\}}$, so that the corresponding set of
transmitted bits $\left\{s_i\right\}_{i \in a}$ must include an
even number of $1$ bits. A particular code is specified by the
set $\mathcal{P}$ of parity checks, each one represented by a
given set of indices~$a$, while the number $|\mathcal{P}|$ of
(independent) parity checks determines the \emph{rate} of the
code ${\rho = 1 - |\mathcal{P}|/N}$. A sequence $\bi{s}$ that
satisfies all the required parity checks ${a \in \mathcal{P}}$
is called a \emph{codeword}. The structure of a code can also
be represented by a bipartite graph, whose two node classes are
associated to bits and parity checks (labelled respectively by
$i$ and $a$), with a link whenever ${i \in a}$ (i.e. when the
bit $i$ is involved in the parity check $a$). Such a graph is
usually called a \emph{Tanner graph}, or a \emph{factor graph}
in more general contexts~\cite{KschnischangFreyLoeliger2001}.
As previously mentioned, for LDPC codes the factor graph is
sparse and locally similar to a tree.

The memoryless BSC flips each bit independently with a given
probability ${x}$, so that the received sequence, which we
shall denote by ${\bi{r} \equiv [r_1,\dots,r_N]}$, may differ
from the transmitted one. In order to perform error-correction
at the receiver, the key quantity to be considered is the
a-posteriori probability, that is, the conditional probability
that the transmitted sequence is $\bi{s}$, given that the
received sequence is $\bi{r}$, which we shall denote as
${p}(\bi{s}|\bi{r})$. It is a standard exercise of Bayesian
inference to show that such a probability can be written as
\begin{equation}
  {p}(\bi{s}|\bi{r})
  \propto
  \chi(\bi{s})
  \left( \frac{{x}}{1-{x}} \right)^{{D}(\bi{s},\bi{r})}
  \, ,
  \label{eq:conditional_probability}
\end{equation}
where $\chi(\bi{s})$ is a characteristic function, taking value
$1$ if $\bi{s}$ is a codeword and $0$ otherwise, and
${D}(\bi{s},\bi{r})$ is the \emph{Hamming distance} (i.e. the
number of different bits) between $\bi{s}$ and~$\bi{r}$. The
proportionality symbol stands for a suitable prefactor
(depending on $\bi{r}$), which ensures the correct
normalization of the conditional probability, namely
\begin{equation}
  \sum_{\bi{s}} {p}(\bi{s}|\bi{r}) = 1
  \, ,
\end{equation}
the sum running over all possible bit sequences of length~$N$.

For algebraic convenience, it is useful to represent each bit
by an Ising-like spin variable ${s_i,r_i = \pm 1}$ (the bit
values ${0,1}$ are mapped respectively to the spin values
${+1,-1}$). As previously mentioned, the a-posteriori
probability can thus be rewritten in the form of a Boltzmann
distribution for an Ising-like model: the transmitted and
received ``spins'' play the role of configuration variables and
external fields, respectively, while the parity checks play the
role of (multispin) interactions. Let us first observe that, in
terms of spin variables, the Hamming distance can be written as
\begin{equation}
  {D}(\bi{s},\bi{r})
  = \sum_{i=1}^{N} \frac{1 - {r}_i {s}_i}{2}
  \, ,
  \label{eq:hamming_distance}
\end{equation}
because the $i$-th term of the summation is equal to $0$ if ${r_i
= s_i}$ and $1$ otherwise. Furthermore, it is useful to define a
function ${V}(\bi{s})$, which counts the number of parity checks
violated by a generic sequence $\bi{s}$. In the spin
representation, such a function can be written as
\begin{equation}
  {V}(\bi{s})
  = \sum_{a \in \mathcal{P}} \frac{1 - \prod_{i \in a} {s}_i}{2}
  \, ,
  \label{eq:violation_function}
\end{equation}
because the spin product ${\prod_{i \in a} s_i}$ takes values $+1$
or $-1$, if the parity check represented by $a$ is satisfied or
not, respectively. With the above definitions, the a-posteriori
probability~\eref{eq:conditional_probability} can be rewritten as
\begin{equation}
  {p}(\bi{s}|\bi{r})
  \propto
  \rme^{-[{D}(\bi{s},\bi{r}) + {J}{V}(\bi{s})]
  \ln \frac{1-{x}}{{x}}}
  \, ,
  \label{eq:conditional_probability_2}
\end{equation}
where we have introduced the fictitious parameter ${J \to
\infty}$, whose role is to set at zero the probability of
non-codewords, that is, of every sequence $\bi{s}$ such that
${V(\bi{s}) > 0}$. Now, making use of
\eref{eq:hamming_distance} and~\eref{eq:violation_function},
and defining the ``Hamiltonian'' function
\begin{equation}
  {H}(\bi{s};\bi{r})
  \triangleq
  - J \sum_{a \in \mathcal{P}} \prod_{i \in a} {s}_i
  - \sum_{i=1}^{N} {r}_i {s}_i
  \, ,
\end{equation}
we can easily write
\begin{equation}
  {D}(\bi{s},\bi{r}) + {J}{V}(\bi{s})
  = \frac{ {H}(\bi{s};\bi{r}) }{ 2 }
  + \mathrm{const}
  \, .
  \label{eq:hamiltonian_plus_const}
\end{equation}
Note that the semicolon in the expression ${H}(\bi{s};\bi{r})$
is meant to distinguish the configuration variables $\bi{s}$
from the parameters (external fields) $\bi{r}$. Note also that
the infinitely large coupling constant $J$ penalizes
(infinitely) the ``excited states'' of each group of spins
involved in a multispin interaction, i.e. it prohibits
parity-check violation. Defining also
\begin{equation}
  \beta
  \triangleq
  \frac{1}{2} \ln \frac{1-{x}}{{x}}
  \, ,
  \label{eq:nishimori_temperature}
\end{equation}
after \eref{eq:conditional_probability_2}
and~\eref{eq:hamiltonian_plus_const} we can finally write a
Boltzmann-like probability
\begin{equation}
  {p}(\bi{s}|\bi{r})
  \propto
  \rme^{ - \beta {H}(\bi{s};\bi{r}) }
  \, ,
\end{equation}
where $\beta$ plays the role of (inverse) temperature. In the
literature of error-correcting codes, the temperature defined by
equation~\eref{eq:nishimori_temperature} is sometimes called the
Nishimori temperature~\cite{Nishimori2001}.

\section{Decoding algorithms}
\label{sec:algorithms}

In order to minimize the \emph{frame error rate} of the decoder
(i.e. the probability that the output sequence ${{\bi{s}^{*}}
\equiv [{s}_1^*,\dots,{s}_N^*]}$ is different from the
transmitted sequence), one has to choose the sequence
${\bi{s}}^{*}$ with the maximum a-posteriori (MAP) probability.
Unfortunately, this criterion requires us to explore all the
codewords, whose number grows exponentially with~$N$, giving
rise to a computationally hard task. An alternative criterion
is to maximize the posterior probability of each single bit,
i.e. the marginal
\begin{equation}
  {p}_i({s}_i|\bi{r})
  = \sum_{\bi{s}|{s}_i} {p}(\bi{s}|\bi{r})
  \, ,
\end{equation}
the sum running over all bit sequences of length $N$ with the
$i$-th bit fixed. The $i$-th output bit is thus chosen as
\begin{equation}
  {{s}_{i}}^{\!\!*}(\bi{r})
  = \arg \max_{{s}_i = \pm 1} {p}_i({s}_i|\bi{r})
  \, .
  \label{eq:bit-map_criterion}
\end{equation}
The latter policy, which is the one adopted in the current
paper, is usually called bit-MAP~\cite{MezardMontanari2009} or
MPM (maximizer of the posterior marginal)~\cite{Nishimori2001}.
It is evident that, in principle, the computational complexity
is still exponential in~$N$, but in this case one can resort to
approximate mean-field methods (such as the Bethe--Peierls
approximation), which provide a direct evaluation of the
marginals, being amenable to a much more efficient numerical
implementation. The marginals can be written in the form of
Boltzmann probabilities for single spins interacting with
effective fields, namely,
\begin{equation}
  {p}_i({s}_i|\bi{r})
  \propto
  \rme^{\beta {h}_i(\bi{r}) {s}_i}
  \, .
  \label{eq:bit_marginal}
\end{equation}
The sign of each effective field ${h}_i$ determines the spin
value ${{s}_i = \pm 1}$ having larger probability, so that the
MPM criterion~\eref{eq:bit-map_criterion} can be easily
rephrased in terms of effective fields. If ${{h}_i = 0}$, both
spin values have equal probability $1/2$, and the output bit is
chosen at random.

\subsection{Bethe--Peierls approximation and belief propagation}

In the Bethe--Peierls approximation scheme, the effective
fields (which turn out to depend on~$\beta$ as well) are
defined implicitly by a set of simultaneous equations, which
also include auxiliary unknowns ${u}_{a \to i}$, called
\emph{messages} in the BP jargon~\cite{MezardMontanari2009},
associated to the links of the factor graph:
\begin{eqnarray}
  {h}_i
  = {r}_i + \sum_{a \ni i} {u}_{a \to i}
  \, ,
  \label{eq:bp_effective_field} \\
  \tanh(\beta {u}_{a \to i})
  = \tanh(\beta J) \prod_{j \in a \setminus i}
  \tanh[\beta ({h}_j - {u}_{a \to j})]
  \, ,
  \label{eq:bp_cavity_bias}
\end{eqnarray}
where the sum runs over all parity checks involving $i$, and
the product runs over all bits involved in the parity check~$a$
except~$i$. Note that by plugging equation
\eref{eq:bp_effective_field} into~\eref{eq:bp_cavity_bias}, one
obtains a set of simultaneous recursive equations for the
messages ${u}_{a \to i}$ alone, where the received signals
${r}_i$ and the noise $\beta$ play the role of parameters. The
iterative self-consistent solution of these equations is an
instance of BP (in fact, it can be regarded as a propagation of
probabilistic information over the factor graph). The resulting
decoding algorithm is described below, the symbol ``$:=$''
denoting the usual assignment statement.

\begin{enumerate}

\item \label{step:initialization} Initialization: for ${i =
    1,\dots,N}$, set ${{h}_i := {r}_i}$ and ${{u}_{a \to i}
    := 0}$ ${\forall a \ni i}$.

\item \label{step:evaluate_decoded_sequence} Evaluate the
    tentative output sequence ${\bi{s}^* \equiv
    [{s}_1^*,\dots,{s}_N^*]}$ according to equations
    \eref{eq:bit-map_criterion} and~\eref{eq:bit_marginal},
    namely
    \begin{equation}
      {{s}_{i}}^{\!\!*} := \cases{
      \sgn {h}_i & if ${h}_i \neq 0$ \\
      \pm 1 \ \mbox{at random} & if ${h}_i = 0$}
      \, .
      \label{eq:decoded_bit}
   \end{equation}
   If ${\bi{s}}^{*}$ is a codeword, terminate, otherwise continue.

\item \label{step:loop_on_checks} For every parity check
    ${a \in \mathcal{P}}$, compute a new estimate of the
    ``outgoing'' messages according to
    equation~\eref{eq:bp_cavity_bias}, namely
    \begin{equation}
      \hat{u}_{a \to i} := \beta^{-1} \tanh^{-1}
      \prod_{j \in a \setminus i}
      \tanh[\beta ({h}_j - {u}_{a \to j})]
      \qquad \forall i \in a \, ,
      \label{eq:message_update_rule}
    \end{equation}
    where a ``hat'' indicates that the updated messages are
    stored in a different memory location. Recall that we
    are interested in the limit ${J \to \infty}$, so that
    the term ${\tanh(\beta J) \to 1}$ can be dropped.

\item \label{step:loop_on_bits} For every bit ${i = 1,\dots,N}$,
recompute the effective field according to
equation~\eref{eq:bp_effective_field}, using the updated messages,
namely
\begin{equation}
  {h}_i := {r}_i + \sum_{a \ni i} \hat{u}_{a \to i}
  \, ,
  \label{eq:bp-algo_effective_field}
\end{equation}
and then ``align'' the messages values, i.e. assign
${{u}_{a \to i} := \hat{u}_{a \to i}}$ ${\forall a \ni i}$.

\item Go back to step~\eref{step:evaluate_decoded_sequence}.

\end{enumerate}
Let us note that, as usual for decoding, the stop test of BP
(step~(\ref{step:evaluate_decoded_sequence})) is not based on
the convergence of the message values, but rather it requires
that the instantaneous values of the effective fields are
mapped, according to equation~\eref{eq:decoded_bit}, to a valid
output sequence (i.e. a codeword).

Possible variants of the above scheme may be addressed to
reduce the numerical complexity or to improve convergence, or
both. As far as the former issue is concerned, the greatest
difficulties arise from the message update
statement~\eref{eq:message_update_rule}, which requires
repeated evaluations of the hyperbolic tangent function (and
its inverse) and a number of multiplications. This equation may
be replaced by the following, much simpler one
\begin{equation}
  \hat{u}_{a \to i}
  :=
  \min_{j \in a \setminus i} |{h}_j - {u}_{a \to j}|
  \prod_{j \in a \setminus i} \sgn({h}_j - {u}_{a \to j})
  \, ,
  \label{eq:message_update_rule_min-sum}
\end{equation}
derived by taking the limit ${\beta\to\infty}$ (zero
temperature). Note that, according to the Nishimori temperature
definition~\eref{eq:nishimori_temperature}, high $\beta$ means
low bit-flip probability ${x}$ (i.e. high signal-to-noise
ratio), which is a typical regime for applications. The
resulting algorithm, usually called
\emph{min--sum}~\cite{MezardMontanari2009,Yedidia2011}, turns
out to be independent of the parameter ($\beta$ or ${x}$) which
characterizes the noise level. As far as convergence is
concerned, one empirically observes that considerable
improvement can be obtained by a sequential update strategy,
whose basic idea is to use updated message values as soon as
they become available~\cite{SharonLitsynGoldberger2004}. With
respect to the previous scheme, such a strategy consists in
replacing steps \eref{step:loop_on_checks}
and~\eref{step:loop_on_bits} with a unique subroutine, which,
for every parity check ${{a} \in \mathcal{P}}$, performs the
message update~\eref{eq:message_update_rule} and then, for each
bit ${i \in a}$, recomputes ${h}_i$ according
to~\eref{eq:bp-algo_effective_field}\footnote[1]{In this case
the initialization statement ${\hat{u}_{a \to i} := 0}$ must be
added at step~\eref{step:initialization}.} and immediately
executes the alignment statement ${{u}_{a \to i} := \hat{u}_{a
\to i}}$. The best results are obtained if the order in which
the parity checks are processed is changed for each iteration
and taken at random~\cite{BraunsteinMezardZecchina2005}. In the
following, we shall denote the latter strategy as \emph{random
sequential} (RS) update and, accordingly, the BP algorithm
equipped with this strategy will be tagged as RSBP.

\subsection{Probability damping}

Let us now observe that the effective-field update
statement~\eref{eq:bp-algo_effective_field} can be replaced
(without introducing any actual change) by
\begin{equation}
  h_i := h_i + \Delta_i
  \, ,
  \label{eq:h_plus_delta}
\end{equation}
where the difference term $\Delta_i$ can be computed using
either of the following two formulae:
\begin{eqnarray}
  \Delta_i & := r_i + \sum_{a \ni i} \hat{u}_{a \to i} - h_i
  \, ,
  \label{eq:delta1} \\
  \Delta_i & := \sum_{a \ni i} (\hat{u}_{a \to i} - u_{a \to i})
  \, .
  \label{eq:delta2}
\end{eqnarray}
Our proposal is to replace equation~\eref{eq:h_plus_delta} with
the following one
\begin{equation}
  h_i := h_i + (1-\gamma) \Delta_i
  \, ,
\end{equation}
where the differential term is attenuated by the factor
${1-\gamma}$, and ${\gamma \in [0,1)}$ is the adjustable
\emph{damping parameter}. For ${\gamma = 0}$ we get back an
ordinary BP, while increasing $\gamma$ values decreases the
amplitude of the difference term, and progressively slows down
the algorithm dynamics. Note that for ${\gamma > 0}$ the use of
equations \eref{eq:delta1} or \eref{eq:delta2} is no longer
equivalent. In particular, using \eref{eq:delta1}, the update
statement reads
\begin{equation}
  {h}_i := (1-\gamma) \biggl( {r}_i + \sum_{a \ni i} \hat{u}_{a \to i} \biggr) + \gamma {h}_i
  \, ,
  \label{eq:pdbp-algo_effective_field}
\end{equation}
namely, each updated field turns out to be a convex linear
combination of the ``old'' value with the ``new'' value
computed by the ordinary BP
rule~\eref{eq:bp-algo_effective_field}. As a consequence, the
resulting algorithm turns out to be exactly equivalent to the
PDBP algorithm, proposed in~\cite{Pretti2005}. The use of
\eref{eq:delta2} gives rise to a different algorithm (from now
on tagged as PD$'$BP), which, in spite of the apparent
similarity with PDBP, turns out to exhibit a considerably
different behaviour. To conclude this section, let us note that
both PDBP and PD$'$BP can be implemented either with the
``exact'' message update rule~\eref{eq:message_update_rule} or
with the simplified (min--sum)
rule~\eref{eq:message_update_rule_min-sum}. For simplicity, in
the following analysis we will consider only the latter case.

\section{Numerical results}
\label{sec:results}

In this section we shall investigate the performance of the
proposed decoding algorithms (PDBP and PD$'$BP), compared to
algorithms using ordinary update equations (BP and RSBP) and
compared to the recent DMBP
algorithm~\cite{YedidiaWangDraper2011}. Let us stress the fact
that all these algorithms, except DMBP, preserve BP fixed
points (i.e. stationary points of the Bethe free energy), at
odds with other well-known message-passing schemes (for
example, so-called \emph{tree-reweighted belief
propagation}~\cite{WainwrightJaakkolaWillsky2002} and
\emph{fractional belief
propagation}~\cite{WiegerinckHeskes2003}), designed to compute
stationary points of modified free energy functionals.
Conversely, DMBP perturbs BP fixed points without reference to
any variational functional, but rather through a direct
modification of the update equations, inspired by the so-called
\emph{divide-and-concur} strategy~\cite{GravelElser2008}. As
far as message scheduling is concerned, let us recall that all
of the algorithms analyzed below (except RSBP) implement a
parallel update scheme, since our main focus is on achieving
practical decoding algorithms, which might be more easily
implemented in hardware.

We shall consider a specific code (with block length ${N =
1057}$ and rate ${\rho \approx 0.77}$) available from MacKay's
repository~\cite{MacKay_code}, and already used
in~\cite{YedidiaWangDraper2011} as a test bed for DMBP. This
code is (almost) a \emph{regular code}, since every bit is
involved in the same number of parity checks ${{n}=3}$, while
every parity check involves the same number of bits ${{m}=13}$
(except one check involving $12$~bits).

The different algorithms will be characterized primarily in
terms of their \emph{frame error rate}. In this respect, let us
recall that a decoding failure may occur either when the
algorithm does not find a codeword within the prescribed
maximum number of iterations, or when it finds a codeword
different from the transmitted one. These two kinds of event
are denoted respectively as \emph{detected} or
\emph{undetected} errors.\footnote[2]{In principle, detected
errors may be further divided into two categories, namely,
actual non-convergence or convergence to a non-codeword. We do
not distinguish these two events because, as previously
mentioned, the stop test of our algorithm simply checks whether
the current values of the effective fields correspond to a
codeword, so that convergence to a non-codeword is equivalent
to non-convergence.} The BSC behaviour for each sequence
transmission is described by an \emph{error pattern}
${[r_1s_1,\dots,r_Ns_N]}$, which is (in the spin
representation) an array containing $+1$ and $-1$ values,
respectively denoting bits that have been received correctly or
incorrectly. In the following, the number of $-1$ values in the
error pattern, i.e. the Hamming distance ${D}(\bi{s},\bi{r})$,
will be shortly denoted as the \emph{weight} of the error
pattern itself. Note that the transmitted sequence $\bi{s}$ is
irrelevant to the decoder behaviour, so that simulations can be
performed with a unique~$\bi{s}$ (for instance ${\bi{s} =
[+1,\dots,+1]}$). For each simulation, the number of sampled
error patterns is adjusted to obtain a significant number
(fixed at $300$) of decoding failures.

\subsection{Error floor}

The frame error rate on a BSC with bit-flip probability ${x}$
can be written as
\begin{equation}
  {P}({x}) = \sum_{{d}={d}_\circ}^{N}
  \left.\mathcal{N}_{d}\right. {x}^{d} (1-{x})^{N-{d}}
  \, ,
\end{equation}
where $\mathcal{N}_{d}$ is the number of error patterns of
weight ${d}$ that cause a decoding failure, and ${d}_\circ$ is
the minimum value of ${d}$ such that
${{\mathcal{N}_{{d}}}\neq{0}}$. The coefficients
$\mathcal{N}_{d}$ (and therefore also ${d}_\circ$) depend on
all the parameters that characterize the decoding algorithm,
namely, the maximum number of allowed iterations $\nu$, the
damping parameter $\gamma$ (if any), and (in principle) the
bit-flip probability~${x}$. As mentioned in the previous
section, the last parameter is actually irrelevant for
min--sum-like algorithms, which implies that the frame error
rate ${P}(x)$ is simply a polynomial of degree $N$ (with no
terms of degree less than ${d}_\circ$). We can thus write
\begin{equation}
  {P}({x}) = \sum_{{k}={d}_\circ}^{N}
  \left.\tilde{\mathcal{N}}_{k}\right. {x}^{k}
  \, ,
\end{equation}
where the relationships between the two sets of coefficients
$\tilde{\mathcal{N}}_{k}$ and ${\mathcal{N}}_{d}$ are as
follows:
\begin{eqnarray}
  \frac{\tilde{\mathcal{N}}_{k}}{{{N}\choose{k}}}
  & = \sum_{{d}={d}_\circ}^{k}
  {{k}\choose{d}} {(-1)}^{{k}-{d}} \frac{{\mathcal{N}}_{d}}{{{N}\choose{d}}}
  & \qquad {k}={d}_\circ,\dots,{N}
  \, ,
  \label{eq:n2ntilde} \\
  \frac{\mathcal{N}_{d}}{{{N}\choose{d}}}
  & = \sum_{{k}={d}_\circ}^{d}
  {{d}\choose{k}} \frac{\tilde{\mathcal{N}}_{k}}{{{N}\choose{k}}}
  & \qquad {d}={d}_\circ,\dots,{N}
  \, .
  \label{eq:ntilde2n}
\end{eqnarray}
In particular, we have
${\tilde{\mathcal{N}}_{{d}_\circ}={\mathcal{N}}_{{d}_\circ}}$,
which makes it evident that the asymptotic behaviour of the
frame error rate for low ${x}$ (i.e. the error floor) is
governed by the error patterns of minimum weight (that cause
decoding failure). In formulae:
\begin{equation}
  {P}({x}) = {\mathcal{N}}_{{d}_\circ} {x}^{{d}_\circ}
  + o({x}^{{d}_\circ})
  \, .
\end{equation}
In general, to compute the coefficients
$\tilde{\mathcal{N}}_{{d}_\circ},\tilde{\mathcal{N}}_{{d}_\circ+1},\dots,\tilde{\mathcal{N}}_{\ell}$
(that is, the $\ell$-th order Taylor approximation), one needs
to know
${\mathcal{N}}_{{d}_\circ},{\mathcal{N}}_{{d}_\circ+1},\dots,{\mathcal{N}}_{\ell}$
(that is, one has to study the decoder behaviour with error
patterns of weight up to $\ell$). In the following, we shall
describe several cases in which ${{d}_\circ={3}}$, and we shall
determine the $5$-th order Taylor approximation. Thanks to the
relatively small size of the code, it is possible to analyze
exhaustively the decoder behaviour under all possible error
patterns of weight up to ${{d}=3}$, and therefore to determine
$\mathcal{N}_{3}$ exactly. Conversely, $\mathcal{N}_{4}$ and
$\mathcal{N}_{5}$ can be determined approximately by
simulations, sampling error patterns of given weight
${{d}=4,5}$. Note that such simulations evaluate the fraction
(rather than the total number) of error patterns that cause
decoding failure, i.e. the quantity
${\mathcal{N}}_{d}\left/{{N}\choose{d}}\right.$, which directly
appears in the conversion formulae \eref{eq:n2ntilde}
and~\eref{eq:ntilde2n}.

\subsection{Ordinary BP algorithms}

\begin{figure}[t!]
  \flushright \resizebox{165mm}{!}{\includegraphics*{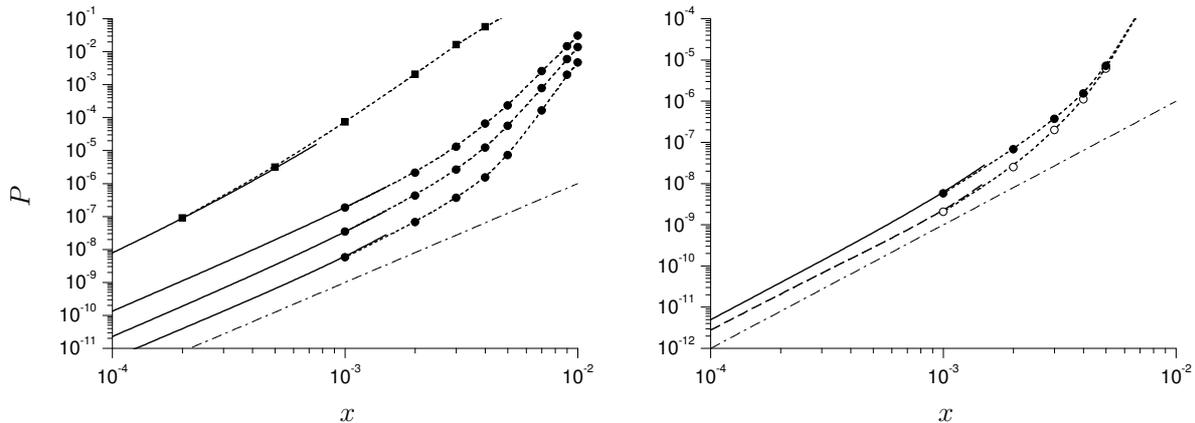}}
  \caption
  {
    Frame error rate ${P}({x})$ for BP (squares) and RSBP (circles),
    with $\nu=20,50,300$ (left panel)
    or $\nu=300$ (right panel).
    Solid and hollow symbols denote respectively the total error rate
    and the one restricted to detected errors.
    Solid and dashed lines represent the 5-th order Taylor approximations,
    respectively. Dotted lines are an eye-guide.
    Thin dash--dotted lines represent ${x}^{3}$.
  }
  \label{fig:min-sum}
\end{figure}

In order to make a better assessment of the improvements that
can be achieved using the modified decoding algorithms (PDBP
and PD$'$BP), we first investigated the behaviour of the
ordinary min--sum algorithm (BP) and of its variant with a
random sequential update (RSBP). The results in terms of frame
error rate ${P}({x})$ are reported in figure~\ref{fig:min-sum},
for different values of the maximum number of allowed
iterations $\nu$. It is noticeable that the performance of BP
is practically insensitive to $\nu$, in the range of values
investigated (the different plots are indistinguishable at the
scale of the figure). RSBP yields remarkably better
performance, which, in addition, can be improved by increasing
$\nu$. This effect suggests that the RS update strategy is able
to eliminate a large number of trapping sets. Nevertheless, in
general, the effect is not quantitatively as relevant as that
observed for the code under consideration, and in any case it
is not of great importance from a technical point of view,
because the RS scheduling cannot be easily parallelized.

\begin{figure}[t!]
  \flushright \resizebox{165mm}{!}{\includegraphics*{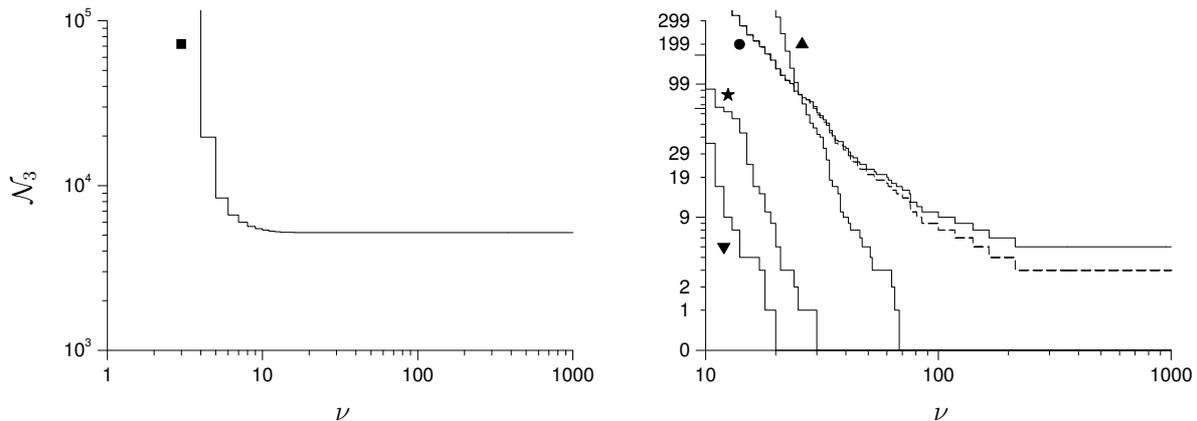}}
  \caption
  {
    Number $\mathcal{N}_{3}$ of error patterns of weight $3$
    that cause decoding failure, generic (solid lines) or detectable (dashed lines),
    as a function of the maximum number of allowed iterations $\nu$,
    for the following algorithms (each one tagged by a symbol):
    BP (square), RSBP (circle),
    PDBP with ${\gamma=0.83}$ (up-triangle),
    PD$'$BP with ${\gamma=0.35}$ (down-triangle),
    DMBP with ${Z=0.35}$ (star)~\cite{YedidiaWangDraper2011}.
    Note that, in principle, the function $\mathcal{N}_{d}(\nu)$
    is defined only for integer values of its argument,
    but for graphical reasons we report the graph of
    $\mathcal{N}_{d}(\lfloor\cdot\rfloor)$
    (which is a right-continuous function), ``welding'' the discontinuities.
  }
  \label{fig:nbiterrric=3}
\end{figure}

We can observe that ${P}(x)$ remains asymptotically
proportional to ${x}^{3}$ for ${{x}\to{0}}$, which means,
according to the above discussion, that we always have
${{d}_\circ={3}}$ (so that we have been able to perform the
exhaustive analysis). Very low values of $\nu$ suffice to
obtain ${{\mathcal{N}_{1}}={0}}$ and ${{\mathcal{N}_{2}}={0}}$,
while the behaviour of ${\mathcal{N}_{3}}$ as a function of
$\nu$ is reported in figure~\ref{fig:nbiterrric=3}. It turns
out that BP reaches a plateau value ${\mathcal{N}_{3}=5180}$ at
${\nu=21}$, and there is no effect of increasing the maximum
number of allowed iterations, at least up to ${\nu=10^6}$. For
RSBP the plateau value is ${\mathcal{N}_{3}=5}$, reached at
${\nu=214}$, which confirms that the number of iterations is
much more effective in this case. Nevertheless, not even RSBP
is able to yield ${\mathcal{N}_{3}=0}$ (i.e. to provide correct
decoding for every error pattern of weight~$3$), due in
particular to the onset of a few undetected errors (the dashed
line in figure~\ref{fig:nbiterrric=3} denotes \emph{detected}
errors), which do not occur for parallel BP. This effect is
probably related to the randomized nature of RSBP, which
increases its ability to explore the configuration space, and
this, on the other hand, is likely to be the same effect that
enables it to escape most trapping sets. The relevance of
undetected errors to the frame error rate (and in particular to
the error floor) of RSBP can be appreciated from the right
panel of figure~\ref{fig:min-sum}. Incidentally, let us observe
that for the code under investigation, the minimum Hamming
distance between codewords is $8$, so that it is not possible,
even with an ideal MAP decoder, to achieve an asymptotic slope
larger than $4$ (i.e. ${\mathcal{N}_{4}=0}$), because a
received sequence with $4$ errors can be at equal distance from
more than one codeword. The right panel of
figure~\ref{fig:nbiterrric=3} shows that the modified
algorithms (namely, PDBP, PD$'$BP and DMBP, with suitable
parameter choices) can achieve the best possible result (i.e.
${\mathcal{N}_{3}=0}$, according to the above argument) for a
relatively low number of iterations, as will be detailed in the
next section.

\subsection{Probability-damping algorithms}

\begin{figure}[t!]
  \flushright \resizebox{165mm}{!}{\includegraphics*{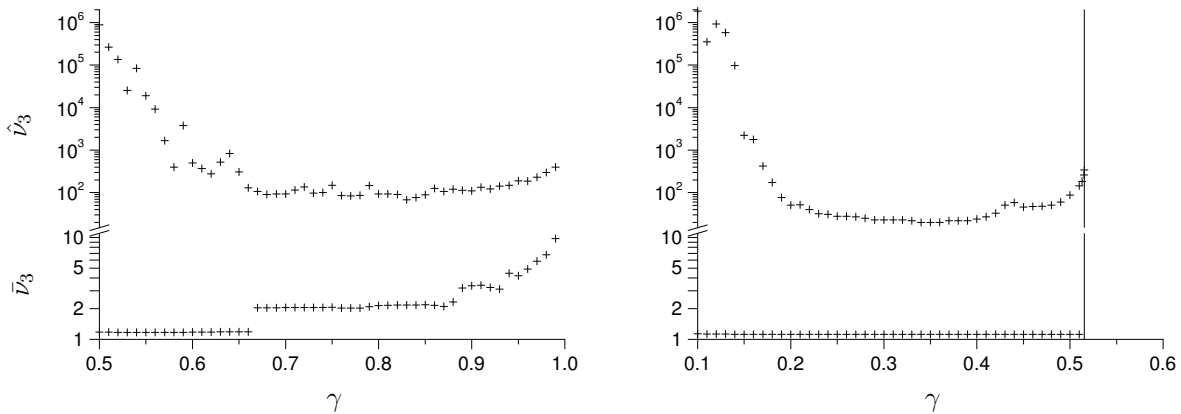}}
  \caption
  {
    Average number $\bar{\nu}_{3}$ and maximum number $\hat{\nu}_{3}$
    of iterations needed to obtain convergence for all error patterns of weight~$3$,
    for PDBP (left panel) and PD$'$BP (right panel),
    as a function of the damping parameter~$\gamma$.
    We have reported data points (cross symbols)
    rather than a continuous interpolating line,
    because the resolution in $\gamma$ that we have been able to
    obtain (with a reasonable computational effort)
    is not sufficient to faithfully represent the fluctuations
    of the dependent variables.
    The solid vertical line marks the threshold $\gamma$ value
    for PD$'$BP (see the text).
  }
  \label{fig:damping}
\end{figure}

Let us now consider PDBP and PD$'$BP in more detail. We have
performed the exhaustive analysis for all error patterns of
weight~$3$ as a function of the damping parameter~$\gamma$ (as
well as ordinary BP, these algorithms yield
${\mathcal{N}_{1}=\mathcal{N}_{2}=0}$ for very low values of
$\nu$). The results are reported in figure~\ref{fig:damping} in
terms of the average and maximum number of iterations needed to
obtain convergence.\footnote[3]{In this case we never encounter
undetected errors, so that \emph{convergence} implies
\emph{correct decoding}.} These quantities, which we
respectively denote by $\bar{\nu}_{3}$ and $\hat{\nu}_{3}$, are
related to the function ${\mathcal{N}_{3}(\nu)}$ (see
figure~\ref{fig:nbiterrric=3}) in a simple way, namely,
considering error patterns of a generic weight ${d}$, we have
\begin{eqnarray}
  \bar{\nu}_{d} & = \frac{1}{{{N}\choose{d}}}
  \sum_{\nu=0}^{\hat{\nu}_{d}-1} \mathcal{N}_{d}(\nu)
  \, , \\
  \hat{\nu}_{d} & = \min \left\{
  \nu \in \mathbb{N} \, | \, \mathcal{N}_{d}(\nu) = 0
  \right\}
  \, .
\end{eqnarray}
Such equations hold if there exists some finite $\nu$ yielding
${\mathcal{N}_{d}(\nu)=0}$, otherwise both $\bar{\nu}_{d}$ and
$\hat{\nu}_{d}$ go to infinity.

We can observe that, for low $\gamma$~values, $\hat{\nu}_{3}$
tends indeed to infinity, which is consistent with the fact
that ordinary BP is not able to obtain convergence for all
error patterns of weight~$3$. For increasing $\gamma$~values,
$\hat{\nu}_{3}$ first exhibits a decreasing trend,
characterized by very large fluctuations, and then a plateau,
with relatively smaller fluctuations. Up to this point, both
PDBP and PD$'$BP show the same qualitative behaviour. Some
quantitative differences appear both in the $\gamma$ value at
which the plateau begins (namely, about $0.65$ for PDBP and
about $0.2$ for PD$'$BP) and in the plateau values of
$\hat{\nu}_{3}$ (roughly $70$--$150$ for PDBP and $20$--$50$
for PD$'$BP). For larger $\gamma$ values, the behaviour of the
two algorithms becomes even qualitatively different. For PDBP,
$\hat{\nu}_{3}$ exhibits just a slight increasing trend, but it
basically remains finite up to ${\gamma\lesssim{1}}$.
Conversely, for PD$'$BP, there is a certain threshold value of
$\gamma$ (slightly above ${0.5}$), such that $\hat{\nu}_{3}$
first undergoes a very sharp increase, and then becomes
practically infinite (in fact we verify that it is larger than
${10}^{7}$). These empirical observations suggest that PDBP and
PD$'$BP, in spite of their similarity, have in fact two very
different dynamical behaviours, namely, even though both
algorithms are able, at the first stage, to eliminate all
possible trapping sets associated to low-weight error patterns,
PD$'$BP appears to introduce new trapping sets upon increasing
$\gamma$. On the contrary, the worsening performance of PDBP
for large growing $\gamma$ values is likely to be ascribed only
to the fact that the algorithm dynamics get slower and slower
in this regime. This conclusion is supported by the fact that,
in this regime, even the average number of iterations
$\bar{\nu}_{3}$ gets larger and larger. It is also noticeable
that, in the low $\gamma$ regime, in which the \emph{maximum}
number of iterations $\hat{\nu}_{3}$ is still very large, the
\emph{average} number of iterations $\bar{\nu}_{3}$ is just
slightly larger than $1$, which means that for most error
patterns of weight~$3$, PDBP converges in $1$~iteration (or a
little more), while there is a very small fraction of
instances, which require a huge number of iterations to
converge. Conversely, at the beginning of the plateau region
of~$\hat{\nu}_{3}$ (precisely at ${\gamma=2/3}$),
$\bar{\nu}_{3}$ exhibits an abrupt jump to slightly above~$2$
(meaning that most instances converge in $2$~iterations), and
in particular one can observe that, for ${\gamma \geq 2/3}$,
absolutely no instances converge in $1$~iteration. The latter
fact can be easily rationalized, taking into account that, at
the first iteration, we have ${{h}_i = {r}_i}$ and ${{u}_{a \to
i} = 0}$. As a consequence, the effective field update
statement~\eref{eq:pdbp-algo_effective_field} becomes
\begin{equation}
  {h}_i := {r}_i + (1-\gamma) \sum_{a \ni i} \hat{u}_{a \to i}
  \, ,
\end{equation}
with ${|\hat{u}_{a \to i}| = 1}$, due to the message update
statement~\eref{eq:message_update_rule_min-sum}. Now, since the
number of summed messages in the above equation is ${n}$ (with
${{n}=3}$ for the current code), if
\begin{equation}
  \gamma \geq 1 - \frac{1}{{n}}
  \, ,
\end{equation}
there is no chance to correct an incorrect sign of ${r}_i$ at
the first iteration. Further similar ``phase transitions'' can
be detected in the large~$\gamma$ regime, where one can
progressively observe no instances converging within
${2,3,4,\dots}$ iterations. As far as PD$'$BP is concerned, the
behaviour of~$\bar{\nu}_{3}$ is completely different, namely,
$\bar{\nu}_{3}$~stays very close to~$1$ in the whole region
where $\hat{\nu}_{3}$ remains finite. This is of course a good
fact for the decoder performance. Nevertheless, the presence of
a region in which $\hat{\nu}_{3}$ becomes infinite at high
$\gamma$, as described above, suggests that the damping
mechanism introduced by PD$'$BP might be less robust, with
respect to PDBP, and that in any case, it requires a finer
tuning of the damping parameter.

By the way, let us recall that in the framework of the
interpretation of PDBP as an approximation of a double-loop
algorithm~\cite{HeskesAlbersKappen2003}, the damping parameter
can be related to a set of \emph{allocation coefficients},
which are associated to the edges of the factor graph and allow
one to define a sufficient condition for
convergence~\cite{Pretti2005}. Assuming that for a regular
factor graph (i.e. a regular code) all the allocation
coefficients are equal, one can map the aforementioned
condition to the following simple inequality:
\begin{equation}
  \gamma \geq 1 - \frac{1}{{n}} \left( 1 - \frac{1}{{m}} \right)^{-1}
  \, .
  \label{eq:convergence_condition}
\end{equation}
The latter can no longer be proved to be a sufficient condition
for convergence of PDBP, but we believe that it can be used as
a ``rule of thumb'' to identify a range of values of the
damping parameter for which the algorithm works best. Note
that, for ${{n}=3}$ and ${{m}=13}$,
condition~\eref{eq:convergence_condition} reads ${\gamma \geq
0.63\overline{8}}$, which, quite surprisingly, roughly
corresponds to the plateau region of $\hat{\nu}_{3}$, described
above (see the left panel of figure~\ref{fig:damping}). A
similar rule is not available for PD$'$BP, since we do not have
an analogous heuristic interpretation for the latter algorithm.

As far as DMBP is concerned, one can observe some similarities
to PD$'$BP, with an even higher sensitivity to the tuning
parameter~$Z$ (see \cite{YedidiaWangDraper2011} for the precise
definition). In particular, $\hat{\nu}_{3}$ remains finite
(with values roughly in the range $30$--$70$) only for ${0.293
\lesssim Z \lesssim 0.433}$, the latter interval being
delimited on both sides by ``sharp thresholds'', like the one
displayed by PD$'$BP. In this interval, the average number of
iterations $\bar{\nu}_{3}$ is almost constant and slightly less
than $2$. One can also verify that the high-$Z$ threshold is
characterized by the onset of a real non-convergence (in
principle we can only state that $\hat{\nu}_{3}$ becomes larger
than $10^7$), while the low-$Z$ threshold is characterized by
convergence to a non-codeword. The latter behaviour is likely
to be ascribed to the considerable distortion of DMBP dynamics
with respect to pure BP. In particular, at odds with PDBP and
PD$'$BP, DMBP does not preserve BP fixed points (except for
${Z=0.5}$), as previously mentioned. Just to give an order of
magnitude for the phenomenon of convergence to non-codewords,
we have analyzed all the failure events collected in our
simulations with error-pattern weight up to~$5$. We observed
that, for DMBP, $100\%$ of detected errors are due to
non-codewords, while, conversely, no such event occurs for pure
BP. The percentage is rather large even for RSBP (about $75\%$)
and PD$'$BP (about $55\%$), while, remarkably, PDBP retains the
pure BP behaviour, with precisely $0\%$ non-codewords.

\begin{figure}[t]
  \flushright \resizebox{165mm}{!}{\includegraphics*{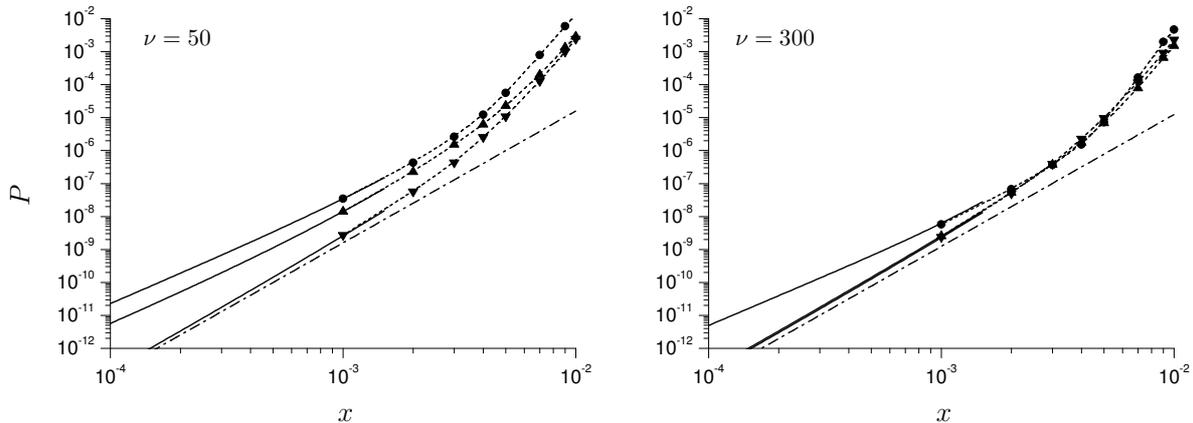}}
  \caption
  {
    Frame error rate ${P}({x})$  for RSBP (circles),
    PDBP with ${\gamma=0.83}$ (up-triangles)
    and PD$'$BP with ${\gamma=0.35}$ (down-triangles),
    for ${\nu=50}$ (left panel)
    and ${\nu=300}$ (right panel).
    The solid lines represent the 5-th order Taylor approximations.
    The dotted lines are an eye-guide.
    The dash--dotted lines represent the asymptotic behaviour
    (4-th order approximation) for DMBP with ${Z=0.35}$.
  }
  \label{fig:performance}
\end{figure}

In the remainder of this section, we discuss the performance of
PDBP and PD$'$BP in terms of the frame error rate. The damping
parameter values (respectively ${\gamma=0.83}$ and
${\gamma=0.35}$) are chosen to minimize $\hat{\nu}_{3}$,
according to the previous analysis (from which we obtain
respectively ${\hat{\nu}_{3}=68}$ and ${\hat{\nu}_{3}=20}$, see
figure~\ref{fig:damping}). The results, reported in
figure~\ref{fig:performance}, are compared to those of RSBP and
of DMBP with the best-performing value of the tuning parameter
${Z=0.35}$\footnote[7]{The equality of this value with the
optimal $\gamma$~value for PD$'$BP is a mere coincidence,
because the two parameters have in fact a different
meaning.}~\cite{YedidiaWangDraper2011}. The behaviour of all
these cases in terms of ${\mathcal{N}_{3}(\nu)}$ has been
reported in figure~\ref{fig:nbiterrric=3}, showing that PD$'$BP
takes the lowest number of iterations to correct all error
patterns of weight~$3$. For ${\nu={50}}$
(figure~\ref{fig:performance}, left panel), we observe that
PDBP slightly improves the error floor of RSBP, but the
asymptotic slope is still $3$, because, according to
figure~\ref{fig:nbiterrric=3}, we still have
${\mathcal{N}_{3}>0}$ (i.e. the algorithm is not yet able to
correct all weight-$3$ error patterns). Conversely, for PD$'$BP
the asymptotic slope is $4$, because ${\mathcal{N}_{3}=0}$
(still according to figure~\ref{fig:nbiterrric=3}). In these
conditions, the error floor behaviour of PDBP is outperformed
by PD$'$BP, whose frame error rate turns out to be almost
coincident with that of DMBP. To avoid confusion, for the
latter algorithm we have reported only the asymptotic behaviour
(4-th order approximation), because the full graph turns out to
be almost completely overlapped with that of PD$'$£BP. Upon
increasing the maximum number of allowed iterations (in the
right panel of figure~\ref{fig:performance} we report the case
${\nu=300}$), the error rates of PDBP and PD$'$BP become more
and more similar to each other, with asymptotic slope~$4$, and
both are practically equivalent to that of DMBP.

\section{Conclusions and perspectives}
\label{sec:conclusions}

In this paper we have considered a BP algorithm with
over-relaxed dynamics, denoted as probability-damping belief
propagation (PDBP)~\cite{Pretti2005}, and a variant of latter
(PD$'$BP), employed as decoding algorithms for low-density
parity-check (LDPC) codes. In terms of the frame error rate, it
turns out that PD$'$BP matches the performance of
difference-map belief propagation
(DMBP)~\cite{YedidiaWangDraper2011}, which is one of the
best-performing iterative decoding algorithms of the BP type
(and of comparable complexity), proposed in the last few years.
Furthermore PDBP achieves basically an equivalent performance,
but only with a larger number (e.g., ${\sim 100}$) of allowed
iterations. The particular code studied, previously used as a
test bed for DMBP, has a minimum distance of $8$, so that any
decoder starts making errors when $4$ bits have been flipped.
Therefore, we have carefully investigated whether decoders can
successfully correct all error patterns of weight~$3$. It turns
out that BP and RSBP do not successfully correct all such
patterns, while PDBP, PD$'$BP, and DMBP do. In particular,
PD$'$BP required the fewest number of iterations to achieve
guaranteed success, followed by DMBP, and then followed by
PDBP. Since practical decoders must limit the number of
iterations to get an acceptable throughput, this is a
significant result, meaning that PD$'$BP is (at least in this
case) the most practical decoder. We recognize that a weak
point in our investigation is the fact that it is based on the
analysis of a single specific code, characterized by a
specially relevant error floor, and on an exceedingly simple
model of a noisy channel (BSC), so it is not clear whether our
claims may hold in a more general setting. Nevertheless,
preliminary (unpublished) results of further simulations
suggest that most of the observations and arguments developed
in this article can be extended to a wider variety of LDPC
codes, and that considerable performance improvements can be
obtained even with a more realistic (e.g. Gaussian) channel
model.

One original feature of PDBP and PD$'$BP, compared to different
BP-inspired decoding
algorithms~\cite{ChenDholakiaEleftheriouFossorierHu2005,ChenTannerJonesLi2005},
including DMBP, is that they preserve BP fixed points. As an
important consequence, we expect they can be used for a much
broader spectrum of problems, in particular when it is
important to actually compute the Boltzmann marginal
probabilities (in the Bethe--Peierls approximation), and
convergence difficulties with pure BP are
encountered~\cite{SomDattaSrinidhiChockalingamRajan2011,WangLiGaoWang2013}.
An additional feature of PDBP, which we have emphasized in the
text, is that one can regard it as an approximation of a
provably-convergent double-loop
algorithm~\cite{HeskesAlbersKappen2003}. As a consequence, we
expect that its more favourable convergence properties might be
quite robust, compared to other algorithms such as PD$'$BP
itself, whose over-relaxing rule has only been devised by
analogy. In particular, we expect that PDBP might perform even
better than PD$'$BP on codes that have already been optimized
to reduce the incidence of trapping
sets~\cite{HuEleftheriouArnold2005,AsvadiBanihashemiAhmadian-Attari2011}.

Finally, we would like to stress two more facts concerning
PDBP, which in our opinion make it rather appealing from a
practical point of view. First, it provides a heuristic rule to
choose the value of the damping parameter. Secondly, it shows
very low sensitivity to variations of this value, at odds with
PD$'$BP, and especially DMBP. The former fact might be very
important for the design of a real decoder. Furthermore,
looking at the original paper in which PDBP was
proposed~\cite{Pretti2005}, one can argue that, in the case of
irregular factor graphs, the aforementioned rule can be
generalized with non-uniform (i.e. bit-dependent) damping
parameters. We expect that such a generalization might provide
further improvements of the error floor for irregular
codes~\cite{ChenTannerJonesLi2005}. As far as the second fact
is concerned, this may even be very important in a practical
implementation, because it should guarantee that the algorithm
performance is not affected by quantization. Of course, a more
detailed assessment of both of these issues is far beyond the
scope of the present paper, as it would require considerable
extra work, but it will probably be the subject of future
research.

\ack I gratefully acknowledge the useful suggestions and
discussions from and with R Zecchina, A Braunstein, F Kayhan
and G Montorsi.

\end{document}